\documentclass[conference]{IEEEtran}
\IEEEoverridecommandlockouts

\usepackage{cite}
\usepackage{amsmath,amssymb}
\usepackage{graphicx}
\usepackage{booktabs}
\usepackage{url}
\usepackage[table]{xcolor}
\usepackage{siunitx}
\newcommand{\pmx}{\,$\pm$\,}

\begin{document}

\title{One Agent per Zone: Decentralised Active Inference under Partial Observability for\\
Multi-Zone HVAC Control on Edge Hardware}

\author{

\IEEEauthorblockN{Panagiotis Kasnesis}
\IEEEauthorblockA{Waldiez P.C.\\
Maroussi, Greece}
\and
\IEEEauthorblockN{Lazaros Toumanidis}
\IEEEauthorblockA{Waldiez P.C.\\
Maroussi, Greece}
\and
\IEEEauthorblockN{Christos Chatzigeorgiou}
\IEEEauthorblockA{Waldiez P.C.\\
Maroussi, Greece}
\and
\IEEEauthorblockN{Amalia Contiero Syropoulou}
\IEEEauthorblockA{ThinGenious P.C.\\
Maroussi, Greece}
\and
\IEEEauthorblockN{Miguel De Prado}
\IEEEauthorblockA{PRAESC AI\\
Biel, Switzerland}
}

\maketitle

\begin{abstract}
Multi-zone HVAC control is a shared-resource problem: zones on a floor share one fan and cooling coil, so each zone's demand affects its neighbours and the building's peak power. One controller per zone scales well and isolates faults, but each controller then observes only its own zone, not the shared air loop. In this work, we present a decentralised controller that assigns
each zone an independent active-inference agent, which plans by minimising expected free energy and learns its zone dynamics online. The agents exchange no messages, so none depends on peers or a coordinator; each accounts for the unobserved coupling through a latent disturbance and a congestion signal
derived from the metered power of the other floors. Fifteen agents are evaluated in EnergyPlus on a Sinergym office building against three rule-based controllers and three MADDPG fleets, one given the same floor meters. On five noisy weather years unseen during training, the fleet attains the lowest mean building peak (\SI{88.3}{\kilo\watt}, 3 to 17\,\% below the baselines) and staggers the floor peaks better than any baseline, with a coincidence factor (building peak over the sum of the floor peaks) of 0.929 versus 0.936 to 0.994. Compared with a reactive rule at similar comfort (0.912 versus 0.916), it lowers the sum of monthly peaks, on which demand charges are billed, by \SI{33}{\kilo\watt} (4\,\%) in every year, at the cost of about 10\,\% more energy. It costs 9\% less per year than the proactive rule (at lower comfort) and 1\% less than the MADDPG fleet given the same floor meters, in every test year. The whole fleet runs on a single low-cost Raspberry~Pi~5, closing the observation-to-action loop over MQTT in
\SI{442}{\milli\second} at p95, far below the 15-minute control step.
\end{abstract}


\section{Introduction}
\label{sec:introduction}

Heating, ventilation and air conditioning (HVAC) systems account for a large share of the electricity consumed by commercial buildings, and their role is no longer limited to keeping the indoor temperature within a comfort band. Many commercial tariffs now charge for the highest power drawn in each billing month as well as for the energy consumed, and grid operators are increasingly interested in when buildings draw power, not only in how much they consume. As a consequence, a modern building controller has to maintain occupant comfort while also limiting and shaping the demand profile of the building.

For a building served by several air loops, the building peak depends not
only on how large the peak of each loop is, but also on whether these peaks
occur at the same time. This is measured by the coincidence factor, the ratio
of the building peak to the sum of the individual loop peaks: a value of 1
means that all loops peak simultaneously, and lower values mean that their
peaks are staggered in time. Reducing coincidence therefore lowers the
building peak without necessarily reducing the energy used by each loop.

In a multi-zone building, the shared HVAC infrastructure induces a
multi-agent control problem: zones served by the same air handler compete for limited fan and cooling capacity, so the actions of one zone affect the thermal conditions of the others. A single controller that decides all setpoints jointly could account for this coupling, but it scales poorly with the number of zones, is a single point of failure, and has to be redesigned whenever the zoning changes. We therefore consider a decentralised architecture in which each zone is controlled by an independent agent that observes only its own zone, without access to the full building state or
communication with neighbouring agents. Variables such as the supply air conditions and the demand of the other zones still influence the local dynamics, but they are not observable by a zone agent and are therefore treated as latent disturbances.

This partial observability matters for two reasons. First, a zone agent cannot directly distinguish changes caused by its own thermal load from changes induced by the other zones on the same air handler; it has to infer their effect from its local observations. Second, peak demand is a collective phenomenon: the building peak is set when several zones call for cooling at the same time, which no zone can see from its own measurements. Effective decentralised control therefore requires agents that reason about hidden influences and, as we show, a minimal aggregate signal that reveals how busy the rest of the building is.

The dominant learning-based approach to this problem has been multi-agent deep
reinforcement learning (DRL) under centralised training with decentralised execution
(CTDE)~\cite{lowe2017,yu2021}, in which a centralised critic observes the whole building during
training while each deployed actor sees only its own zone. Although this
paradigm works, it inherits the properties of its training regime: the
coupling between zones is absorbed into the network weights rather than being
represented explicitly, and the resulting policy is difficult to inspect at the
moment it matters.

In this paper, we take a different route and assign each zone to an
independent \emph{active-inference} (AIF) agent~\cite{friston2017}. Each agent
maintains beliefs over a small set of discrete latent states, corrects them
with Bayes' rule as new observations arrive, learns its own transition model
online, and selects actions by evaluating how well each candidate action
satisfies its comfort and energy preferences over the next two hours. The
generative model also contains a hidden disturbance factor that is never
observed: when a zone systematically runs warmer or cooler than its model
predicts, the agent's belief about this factor shifts accordingly, which is how
an agent accounts for an air-loop coupling that it cannot measure. In addition,
every agent reads the metered power of the other two air loops as a congestion
signal and becomes more reluctant to draw power when the rest of the building
is already busy. AIF has recently been applied to building
energy management with one agent per building and a supervising
community-level agent~\cite{nazemi2025}; however, to the best of our knowledge,
this is the first work that applies decentralised AIF
at the zone level to multi-zone HVAC control over a shared air loop, and
evaluates it against both rule-based control (RBC) and DRL
controllers on demand-related metrics.

The main contributions of this work are summarised as follows:
\begin{itemize}
\item We present a decentralised AIF controller for shared-resource HVAC
systems, in which 15 zone agents stagger their demand across floors without
exchanging a single message. Evaluated on five noisy weather years that were
not seen during training, the AIF fleet achieves the lowest mean annual peak
(\SI{88.3}{\kilo\watt}) among the nine controllers that we tested and a lower
coincidence factor (0.929) than any baseline.

\item We identify why the fleet coordinates. Our ablations show that two ingredients are needed together: a real-time signal of what the rest of the building is drawing, and agents that respond to it differently from one another. Replacing the live signal with a constant of the same mean, or with the signal of the previous day, removes the effect, which rules out a higher energy price or a memorised daily pattern as explanations. Giving the same meters to a MADDPG fleet also lowers its peak, which supports the conclusion that the live load information, and not active inference alone, is what enables the staggering.
\item We provide a transparent comparison with well-tuned RBCs, reporting the metrics on which simple rules still prevail and
quantifying the demand charge at which the demand savings of the AIF fleet
would compensate for its additional energy use.
\item We show that all 15 agents run concurrently on a Raspberry~Pi~5, with
a p95 observation-to-action latency of \SI{442}{\milli\second} measured over
the full MQTT round trip.
\end{itemize}

The remainder of this paper is organised as follows. Section~\ref{sec:background}
provides the necessary background on AIF, while Section~\ref{sec:related} reviews related
work. Section~\ref{sec:testbed} introduces the simulation testbed, the evaluation
protocol and the controllers. Section~\ref{sec:metrics} defines the evaluation
metrics, Section~\ref{sec:results} presents the obtained results, and
Section~\ref{sec:discussion} discusses their implications and limitations.
Finally, Section~\ref{sec:conclusion} concludes the article. All code, trained models and
benchmark harnesses are released under Apache~2.0 at
\url{https://github.com/waldiez/aif-sinergym}.

\section{Background}
\label{sec:background}

\subsection{Active Inference in Discrete State Spaces}

AIF~\cite{friston2017} describes an agent that holds a
probabilistic generative model of how its observations are produced by hidden
states of the world, and of how these states evolve under its own actions. In
the discrete formulation used in this work, the model is specified by a
likelihood mapping $\mathbf{A}$, which gives the probability of each
observation given the hidden state, a transition model $\mathbf{B}$, which
gives the probability of the next state given the current state and the
selected action, and a preference distribution $\mathbf{C}$ over observations,
which encodes the outcomes that the agent seeks. When the state space is
factored, each factor (e.g., temperature, occupancy or a latent disturbance)
has its own likelihood and transition terms, which keeps the model small
enough to be updated in real time.

At every time step, the agent first infers the posterior over hidden states by
combining its prediction from the previous step with the likelihood of the new
observation. It then evaluates each candidate policy through its expected free
energy, which can be decomposed into a \emph{pragmatic} term, rewarding
predicted outcomes that match the preferences in $\mathbf{C}$, and an
\emph{epistemic} term, rewarding actions that are expected to reduce
uncertainty about the hidden states. As a result, information seeking is part
of the objective itself rather than an exploration heuristic added on top of
it. Finally, the transition model can be learned online by accumulating
Dirichlet counts over observed transitions, so the agent refines its own model
of the plant while it controls it.



\section{Related Work}
\label{sec:related}

Reinforcement learning for HVAC control has been studied extensively on
simulation benchmarks built over EnergyPlus~\cite{crawley2001}, among which
Sinergym~\cite{jimenezraboso2021,campoynieves2025} is the framework adopted in
this work.
Multi-agent formulations typically follow MADDPG~\cite{lowe2017} with the
stabilisation techniques of TD3~\cite{fujimoto2018}, which is also the DRL baseline that we implement.
Single-agent deep RL has been benchmarked extensively on such testbeds.
Campoy-Nieves et al.~\cite{campoynieves2025} compared SAC, TD3 and PPO with
default and RBC in a two-zone data centre, where PPO learned
the most stable policies and saved energy mainly in the intermediate seasons.
Zhan et al.~\cite{zhan2025} compared six algorithms across office and
data-centre buildings, climates and reward functions, and reported energy
reductions of about 9 to 10\,\% over RBC, with an exponential
comfort penalty reducing temperature fluctuations compared with a linear one.
In a national-scale energy setting, Bajrami et al.~\cite{bajrami2025} found
SAC faster and more stable than PPO, whereas PPO produced more assertive
actions. Sample efficiency remains a central obstacle for these methods: Qu
et al.~\cite{qu2025} augmented a SAC heating controller with an LSTM surrogate
and branched rollouts, more than halving its training time, while Kadamala et
al.~\cite{kadamala2025} pre-trained agents by behavioural cloning and
fine-tuned them across buildings with different observation spaces by padding
the input features. All of these works control the building with a single
agent; the coupling between zones that share an air loop, which is central to
our problem, is not modelled explicitly.

AIF has so far been applied mainly to low-dimensional control
problems, and \texttt{pymdp}~\cite{heins2022} provides a reference
implementation for discrete state spaces. Its appeal for building control
stems from two properties: the epistemic value is part of the objective rather
than an exploration rule added afterwards, and the generative model can be
inspected at decision time, which makes the behaviour of each agent easier to
interpret. Applications of AIF to energy systems are still
scarce compared to DRL. Nazemi et al.~\cite{nazemi2025} proposed a dual-layer architecture in
which a continuous AIF agent per building sets the HVAC airflow and
supply-air temperature while inferring occupancy and infiltration as hidden
states, and a discrete AIF agent coordinates storage, photovoltaics and market
transactions at the community level. The building agent matched a
full-information optimisation benchmark and compared favourably with a DDPG
controller. In that work, however, each building is modelled as a single zone
and coordination is achieved hierarchically through the community agent. Our
agents instead operate at the zone level, share a physical air loop, and
coordinate without any supervising agent or message exchange.

\section{Testbed and Controllers}
\label{sec:testbed}

\subsection{Building and Simulation}
We use Sinergym~\cite{jimenezraboso2021,campoynieves2025} driving
EnergyPlus~24.1.0 on the ASHRAE 90.1 OfficeMedium prototype
building~\cite{deru2011} (Denver model,
\SI{4979.6}{\metre\squared}, three floors), but with a custom environment rather than one of
the predefined Sinergym scenarios: we expose per-zone setpoints as actions,
extend the observation space, and add one electricity meter per air loop, as
described below.
In particular, of its 18 thermal zones, 15 are
conditioned (3 cores and 12 perimeters), while the remaining 3 are return
plenums. Each floor is served by one packaged air loop, in which five zones
share one fan, one two-speed cooling coil and one outdoor-air system, while
each zone has its own damper, electric reheat coil and dual-setpoint
thermostat. In addition, we added one electricity meter per air loop, covering
the fan, the cooling coil and the reheat coils of the corresponding floor;
such a meter per air handler is a modest addition in a commercial building.

The observation vector consists of 95 values: time, outdoor conditions, 18
zone temperatures and humidities, 15 occupancy counts, 30 setpoints, the
whole-building HVAC demand and energy, and the three floor meters. The action
vector contains a heating and a cooling setpoint for each conditioned zone (30
values), bounded to \SIrange{15.0}{22.5}{\celsius} for heating and
\SIrange{22.5}{30.0}{\celsius} for cooling, as specified by our experimental setup using the sinergym simulator. It should be noted that these
bounds do not by themselves keep the two setpoints apart, which is why we
report deadband violations for every controller. The simulation timestep is 15
minutes, which results in 35\,040 steps per simulated year.

\subsection{Evaluation Protocol}
\label{sec:protocol}

Every controller is evaluated under three weather conditions. The first one,
denoted as W0, is the deterministic ``mixed'' TMY3 year (New York JFK), and it
is the only year on which the learning controllers are trained or adapted. The
second one, W1, consists of the same year perturbed by Ornstein-Uhlenbeck noise
on dry-bulb temperature ($\sigma = \SI{1.5}{\celsius}$) and humidity, generated
with five different seeds; these five years constitute our main test set. For
a given seed, every controller is exposed to exactly the same weather, which we
verify by comparing the logged outdoor temperatures, so that all comparisons on
W1 can be paired by year. Finally, W2 comprises the hot (Davis-Monthan, Arizona) and cool (Port
Angeles, Washington) TMY3 years provided with Sinergym, which serve as
unseen climates.
All learned controllers are tested with learning switched off, so that no
controller adapts to its test years. All controllers target, and are scored
against, the same seasonal comfort band: \SIrange{20.0}{23.5}{\celsius} in
winter and \SIrange{23.0}{26.0}{\celsius} in summer (1~June to
30~September).

\subsection{Controllers}

All controllers share the same interface: every 15 minutes they receive the
observation vector and return the 30 setpoints. The learning controllers are
deployed in a decentralised manner, with one agent per zone.

\subsubsection{Rule-based controllers}
The two main RBCs are implemented in a single code base and differ in exactly
two behavioural switches, so that the pair isolates the value of anticipation
and structural awareness rather than comparing unrelated heuristics. The
\emph{reactive RBC} applies a seasonal threshold rule on the temperature of
each zone and on an occupancy calendar,
with a \SI{2}{\celsius} setback when the building is unoccupied and without any
anticipation. The \emph{proactive RBC} additionally performs a pre-conditioning
phase between 04:00 and 06:00 when the outdoor temperature is low, and applies
per-zone offsets derived from the core or perimeter position, the orientation
and the floor of each zone. 
The proactive RBC has one setting that controls how hard it cools in summer,
the cooling margin. We run it with margins from \SIrange{0.5}{2.5}{\celsius}:
a small margin gives more comfort at a higher cost, and a large margin the
opposite. This shows the full range of comfort and cost that this rule can
reach (Fig.~\ref{fig:pareto}), so that the AIF fleet is not compared with a
single tuning only.

Moreover, we implemented a \emph{demand-limiting RBC}, which extends the
proactive RBC with two textbook peak-management measures. First, the three
floors start their pre-conditioning 40 minutes apart; second, whenever the
building HVAC demand exceeds 90\,\% of \SI{92.4}{\kilo\watt} (the annual peak
of the proactive RBC), all cooling setpoints are raised and all heating
setpoints are lowered by \SI{1}{\celsius} until the demand falls again. We also
tested other thresholds, a larger shed and a rotating shed across floors, none
of which lowered the peak.

\subsubsection{MADDPG}
As a DRL baseline, we use a multi-agent actor-critic
controller~\cite{lowe2017} with TD3 stabilisers~\cite{fujimoto2018}, trained
with a centralised twin critic and deployed as 15 independent actors. Each
actor receives a 29-dimensional feature vector describing its zone, which
contains time, weather, whole-building HVAC demand and energy, and local zone
features including a fixed zone identity. The actor consists of a single-layer
GRU with 64 hidden units applied over the last eight steps, followed by a small
multilayer perceptron, for a total of about 40k parameters. The networks were
trained on the shaped reward described in Section~\ref{sec:metrics} for 50
simulated years, corresponding to roughly 3.5 million gradient updates. We
evaluate three fleets: a cold start, a start whose replay buffer was seeded
with one year of the reactive RBC, and a cold start whose actors additionally
receive the three floor meters (own floor, other floors and whole building),
so that the latter has access to the same information as the AIF agents.

\subsubsection{Active inference}
Each zone agent maintains a factored discrete generative model with four
factors: the zone temperature $T$ (19 bins, \SI{0.5}{\celsius} wide near the
comfort band), the occupancy $O$ (2 states, derived from the weekday 06:00 to
20:00 calendar), the outdoor temperature $W$ (5 bins), and a hidden disturbance
$H$ (5 bins spanning $\pm\SI{0.8}{\celsius}$) that is never observed. The agent
selects among 25 setpoint pairs, formed by five heating values (15.0 to
\SI{22.5}{\celsius}) and five cooling values (22.5 to \SI{30.0}{\celsius}).

The transition model $\mathbf{B}$ is initialised from a simple physical prior,
in which the HVAC drives the zone towards the commanded band and heat is lost
towards the outdoor temperature at a rate that depends on whether the zone is a
core, a perimeter or located on the top floor. The agent then refines this
model from its own experience by means of Dirichlet count updates. Its
preferences $\mathbf{C}$ reward temperatures inside the comfort band, relaxed
by \SI{4}{\celsius} when the zone is unoccupied, and penalise HVAC effort.

At every time step, the agent updates its beliefs and then scores each of the
25 actions by assuming that the action is held for the next eight steps (two
hours):
\begin{equation}
\begin{aligned}
G_i(a) = {} & \sum_{k=1}^{8} \Big( \mathbb{E}[C(T_k)]
  - \big(w^E_i + \lambda_c\, c_i\big)\,\mathbb{E}[D(T_k,a)] \Big) \\
  & + \beta\, I(T_1;H) - \lambda_{db}\, P_{db}(a),
\end{aligned}
\label{eq:efe}
\end{equation}
where $C$ is the comfort preference, $D$ the expected HVAC effort, $w^E_i$ the
energy weight of agent $i$, $c_i$ its congestion signal, $I(T_1;H)$ the
expected information gain about the hidden disturbance, and $P_{db}$ a penalty
for heating and cooling setpoints that are closer than \SI{2.5}{\celsius}. We
use $\lambda_c = 5$, $\beta = 0.2$ and $\lambda_{db} = 8$. Since $G$ is the
negative of the expected free energy, the agent selects the action with the
highest $G$. Finally, a reactive safety rule takes over whenever a zone is more
than \SI{2}{\celsius} outside its comfort band, which occurred in about
4.4\,\% of the decisions.

\textbf{Coordination without communication.} No agent sends any message to
another agent; nevertheless, two terms of Eq.~\eqref{eq:efe} give rise to
coordinated behaviour. First, each agent reads the power drawn by the
\emph{other two} floors from the meters, divides it by a fixed reference of
\SI{92.4}{\kilo\watt} (the annual peak of the incumbent controller, which a
facility manager can read from the previous year's bill), and smooths the
result into a congestion signal $c_i$. In this way, drawing power becomes more
expensive for an agent when the rest of the building is busy. Second, each
agent adapts its own energy weight $w^E_i$ within $[0,1]$, increasing it while
its zone is comfortable and decreasing it otherwise. Since all agents of a
floor receive the same congestion signal, these heterogeneous weights are what
prevent the five agents from reacting in lockstep.

We refer to this configuration as \emph{AIF (fixed reference)}. For
comparison, we also report a variant, \emph{AIF (running reference)}, in which
each agent divides the load of the other floors by the highest total floor
power that it has observed so far instead of a fixed value. This running maximum is taken over the total power of the three floors and
starts from the first load observed after deployment, which is far below the
eventual peak; it then only grows over time, so during the first weeks of
operation the agents perceive the building as more congested than it actually
is. Moreover, its value depends on the history of each deployment, so the same
model can scale the signal differently depending on how long it has been
running and which loads it has observed. The fixed reference avoids both
issues, and we therefore use it as our main configuration, while the running
reference serves as a sensitivity check.

We implemented the agents from scratch in PyTorch, so that the whole fleet is expressed as a set of batched tensor
operations, which runs on a GPU during simulation and on a CPU at the edge.
Importantly, the batch of 15 agents contains no coupling across its batch
dimension: we verified that each row of the batch produces exactly the same
decisions as a separate single-agent model fed with the same inputs, which is
what allows us to deploy 15 truly independent actors from a single model. Each
fleet adapts for five years on W0 and is then frozen for testing.

\section{Evaluation Metrics}
\label{sec:metrics}

All controllers are scored by the same code against the comfort band of Section~\ref{sec:protocol}, with comfort measured on weekdays between 07:00 and 19:59.

\textbf{Comfort.} The comfort score $\mathrm{CS}_{occ}$ assigns each occupied
zone and step a value of 1 when the temperature lies inside the band, and
$1 - \min(v,1)^2$ when it lies $v$\,\si{\celsius} outside it; these values are
then averaged over zones and steps. In addition, the zone comfort rate (ZCR)
reports the share of zone-steps that lie strictly inside the band.

\textbf{Demand.} We report the annual peak of the building HVAC demand, the sum
of the twelve monthly peaks (which is the quantity actually charged by the
tariff), and the coincidence factor
\begin{equation}
\mathrm{CF} = \frac{\max_t \sum_f P_f(t)}{\sum_f \max_t P_f(t)},
\end{equation}
where $P_f$ denotes the power drawn by floor $f$. As explained in
Section~\ref{sec:introduction}, a CF equal to 1 means that the three floors peak
at the same moment, whereas lower values indicate that their peaks are spread
out in time.

\textbf{Cost.} We adopt a simplified tariff modelled on PG\&E
Schedule~B-19~\cite{pge_b19}, with time-of-use energy prices of 0.22, 0.32
and 0.45~\$/kWh (off-peak, mid-peak and on-peak, respectively) and a single
demand charge of \$26 per kW of monthly peak.

\begin{table*}[t]
\caption{Results on the five noisy test years (W1), mean\,$\pm$\,std. Learned
controllers were trained or adapted on the deterministic year only and are
tested with learning switched off. Lower is better for demand, energy and
cost. Higher is better for comfort.}
\label{tab:w1}
\centering
\footnotesize
\begin{tabular}{lccccccc}
\toprule
Controller & Peak & $\Sigma$ monthly & CF & CS$_{occ}$ & ZCR & Energy & Cost \\
 & (kW) & peaks (kW) & & & (\%) & (MWh) & (k\$) \\
\midrule
\textbf{AIF (fixed reference)} & \textbf{88.3}\pmx2.8 & 765\pmx14 & 0.929\pmx0.030 & 0.912\pmx0.003 & 83.6\pmx0.4 & 117.4\pmx2.7 & 55.4\pmx1.1 \\
AIF (running reference)     & 88.5\pmx3.6 & \textbf{763}\pmx14 & \textbf{0.927}\pmx0.028 & 0.908\pmx0.003 & 82.9\pmx0.5 & 116.7\pmx2.6 & 55.0\pmx1.1 \\
AIF, no congestion signal   & 94.1\pmx2.3 & 784\pmx16 & 0.967\pmx0.015 & 0.938\pmx0.001 & 88.7\pmx0.2 & 121.2\pmx2.8 & 57.8\pmx1.2 \\
RBC reactive                & 93.8\pmx0.5 & 799\pmx14 & 0.974\pmx0.005 & 0.916\pmx0.003 & 85.6\pmx0.5 & \textbf{106.2}\pmx2.9 & \textbf{53.7}\pmx1.2 \\
RBC proactive               & 90.9\pmx3.4 & 767\pmx17 & 0.958\pmx0.014 & \textbf{0.946}\pmx0.001 & \textbf{90.6}\pmx0.1 & 139.2\pmx2.8 & 61.1\pmx1.3 \\
RBC demand-limiting         & 97.2\pmx2.7 & 781\pmx19 & 0.983\pmx0.007 & \textbf{0.946}\pmx0.001 & 90.5\pmx0.1 & 138.0\pmx2.8 & 61.2\pmx1.3 \\
MADDPG cold start           & 106.0\pmx2.1 & 787\pmx5 & 0.994\pmx0.005 & 0.924\pmx0.001 & 83.1\pmx0.3 & 120.3\pmx2.4 & 56.5\pmx0.8 \\
MADDPG warm start (reactive)& 103.5\pmx4.0 & 798\pmx16 & 0.977\pmx0.015 & 0.931\pmx0.002 & 84.3\pmx0.3 & 121.0\pmx2.6 & 57.0\pmx1.2 \\
MADDPG + floor meters       & 97.0\pmx2.0 & 774\pmx15 & 0.936\pmx0.015 & 0.926\pmx0.002 & 83.4\pmx0.4 & 119.6\pmx2.6 & 55.9\pmx1.1 \\
\end{tabular}
\end{table*}

Finally, the MADDPG fleets are trained on a shaped reward that is shared by
all agents and computed at every step $t$ as
\begin{equation}
\begin{aligned}
r_t = {} & -\Big( w_E\, m_t\, \rho(P_t) + w_C\, \kappa_t\, \phi_t \\
        & \quad + w_{DB}\, \delta_t + w_P\, \rho(\Delta P^{pk}_t) \Big)
          + w_M\, \mu_t\, \psi_t ,
\end{aligned}
\label{eq:reward}
\end{equation}
where $P_t$ is the building HVAC demand and
$\rho(x) = \min\!\big(\log(1 + x/P_{max}) / \log 2,\; 2\big)$ is a logarithmic
normalisation with $P_{max} = \SI{300}{\kilo\watt}$. The electricity term is
scaled by the time-of-use multiplier $m_t$, which equals 1, 1.5 and 3 in the
off-peak, mid-peak and on-peak periods, respectively, and is increased by a
further 10\,\% when the building is unoccupied. The comfort penalty $\phi_t$
averages over the zones the band violation $v_z$ of each zone, normalised as
$\min(v_z/s, 1)$; during occupied hours $s = 4$ and violations in the
direction of the seasonal load (too warm in summer, too cold in winter) are
weighted by 1.3, while before occupancy $s = 5$ and after occupancy the band
is additionally widened by \SI{3}{\celsius}. The occupancy weight $\kappa_t$
equals 1 during occupied hours (07:00 to 19:59 on weekdays), 0.4 before
occupancy (05:00 to 07:00), 0.15 after occupancy (20:00 to 21:59) and 0
otherwise. The deadband penalty
$\delta_t = \frac{1}{15}\sum_z \min\!\big(\max(h_z - c_z + 2, 0)/3,\; 1\big)$
penalises heating and cooling setpoints $h_z$ and $c_z$ that are closer than
\SI{2}{\celsius}, and the peak term penalises the increase
$\Delta P^{pk}_t = \max(P_t - \hat{P}_t, 0)$ over the running monthly peak
$\hat{P}_t$, so that it is non-zero only when a new monthly peak is set.
Finally, the midpoint bonus $\psi_t$ averages
$\max(1 - |T_z - T_{mid}|/2.5,\; 0)$ over the zones, where $T_{mid}$ is the
middle of the comfort band, with $\mu_t$ equal to 1 during occupied hours, 0.5
before occupancy (with a scale of 4 instead of 2.5) and 0 otherwise. We use
$w_E = 0.14$, $w_C = 0.40$, $w_{DB} = 0.05$, $w_P = 0.17$ and $w_M = 0.10$.

\section{Results}
\label{sec:results}

\subsection{Main Comparison on the Noisy Test Years}

Table~\ref{tab:w1} presents the results obtained on the five noisy test years.
Since every controller was exposed to the same weather in each year, we also
compare the controllers year by year in Table~\ref{tab:paired}, reporting the
mean paired difference together with the number of years in which the AIF fleet
performed better.

\begin{table}[t]
\caption{Paired differences, AIF (fixed reference) minus the other controller, mean
over the five W1 years. The count shows in how many years the AIF fleet was
better.}
\label{tab:paired}
\centering
\footnotesize
\setlength{\tabcolsep}{3pt}
\begin{tabular}{lccc}
\toprule
 & RBC & RBC & MADDPG \\
 & reactive & proactive & + meters \\
\midrule
Annual peak (kW)      & $-5.5$ (5/5) & $-2.6$ (3/5) & $-8.7$ (5/5) \\
$\Sigma$ monthly peaks & $-33$ (5/5) & $-2$ (3/5) & $-9$ (3/5) \\
CF                    & $-0.045$ (4/5) & $-0.029$ (4/5) & $-0.007$ (4/5) \\
CS$_{occ}$            & $-0.004$ (0/5) & $-0.034$ (0/5) & $-0.015$ (0/5) \\
Energy (MWh)          & $+11.2$ (0/5) & $-21.8$ (5/5) & $-2.1$ (5/5) \\
Total cost (k\$)      & $+1.6$ (0/5) & $-5.7$ (5/5) & $-0.5$ (5/5) \\
\bottomrule
\end{tabular}
\end{table}

\textbf{Peak demand and coincidence.} The AIF fleet achieves the lowest mean
annual peak (\SI{88.3}{\kilo\watt}, against 90.9 to \SI{106.0}{\kilo\watt} for
the baselines) and a lower coincidence factor than any baseline (0.929). Its
peak is
lower than that of the reactive RBC and of every MADDPG fleet in every single
test year. Against the proactive RBC, however, the annual peak is lower in only
three of the five years, so we do not claim a robust peak advantage over that
controller.

\textbf{Comfort and cost.} The two RBCs span the range of comfort and cost, and
the AIF fleet lies between them. Together with the demand-limiting RBC, the
proactive RBC is both the most comfortable controller ($\mathrm{CS}_{occ}$ of
0.946) and the most expensive one (about \$61k), whereas the reactive RBC is the cheapest (\$53.7k) and among the
least comfortable. The AIF fleet costs \$5.7k less than the proactive RBC in
every year, at the price of lower comfort. Compared with the reactive RBC, it
reaches a similar comfort level (0.912 against 0.916) with a markedly lower and
better-staggered demand, but it consumes about 10\,\% more energy; under the
evaluated tariff, the demand savings do not compensate for this additional
energy, so the AIF fleet costs \$1.6k more per year, as discussed further in
Section~\ref{sec:discussion}.

\begin{figure*}[t]
\centering
\includegraphics[width=0.9\textwidth]{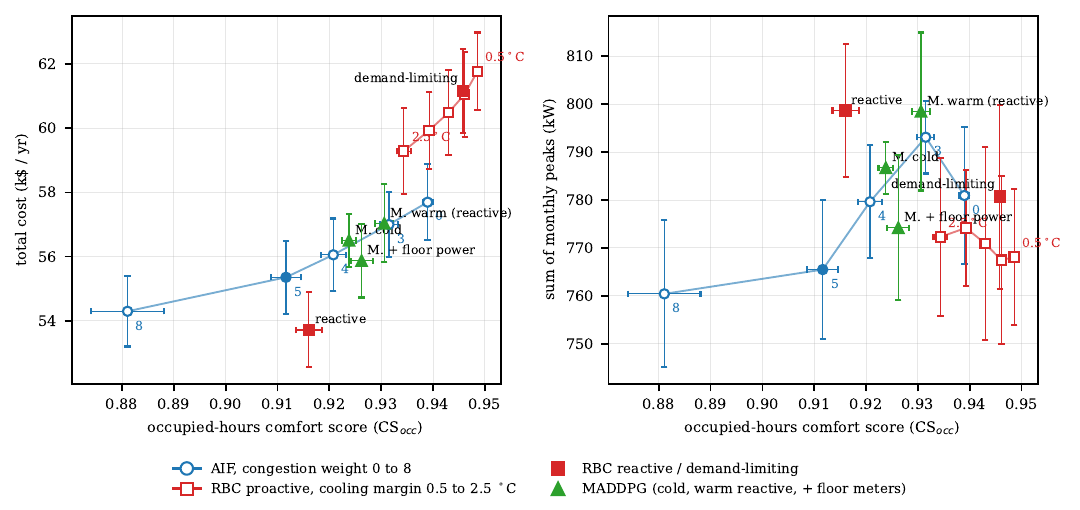}
\caption{Comfort against total cost (left) and against the sum of monthly
peaks (right) on the five noisy test years (mean\,$\pm$\,std over years).
Blue: AIF with adaptive energy weights and congestion weights of 0, 3, 4, 5
and 8 (numbers next to the points); the filled point is AIF (fixed reference)
with weight 5, while the other points use the running reference. Red: the
proactive RBC with summer cooling margins from 0.5 to \SI{2.5}{\celsius},
together with the reactive and demand-limiting RBCs (filled squares). Green
triangles: MADDPG fleets.}
\label{fig:pareto}
\end{figure*}

\textbf{The cooling-margin sweep.} Fig.~\ref{fig:pareto} summarises the
comfort, cost and demand trade-offs of all controllers. Changing the summer
cooling margin of the proactive RBC trades comfort for cost, but its peak
remains between 91 and \SI{93}{\kilo\watt} on the noisy years. Interestingly,
on the deterministic year one setting (\SI{2.5}{\celsius}) happened to miss the
design-day peak and reached \SI{82.8}{\kilo\watt}, whereas on the noisy years
the same setting averaged \SI{92.8}{\kilo\watt}. This observation illustrates
the risk of judging controllers on a single year: an annual peak is a single
15-minute value, and one favourable day can change it substantially.

\textbf{Demand limiting.} The demand-limiting RBC left the peak unchanged on the
deterministic year and even raised it on the noisy years
(\SI{97.2}{\kilo\watt}, against \SI{90.9}{\kilo\watt} for the plain proactive
RBC). This can be explained by the fact that the annual peak occurs on hot
afternoons, when the zones are already above their setpoints, so raising the
setpoint by one or two degrees does not reduce the cooling call; moreover, the
rule reacts only after the demand has already spiked. A tighter limit even
raised the peak further, most likely through a rebound when the shed was
released.

\begin{figure}[t]
\centering
\includegraphics[width=\columnwidth]{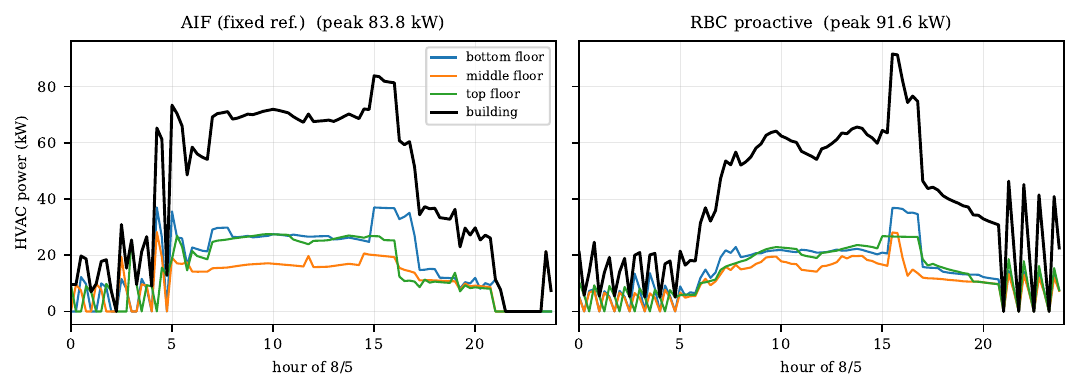}
\caption{HVAC power per floor on 5 August of the deterministic year. Left:
AIF fleet (last adaptation year). Right: proactive RBC.}
\label{fig:peakday}
\end{figure}

\textbf{Behaviour on the design day.} Fig.~\ref{fig:peakday} shows the power
drawn by each floor on 5 August, the day of the annual peak for most
controllers. Under the proactive RBC, all three floors increase their demand
together at about 15:30, and the building reaches \SI{91.6}{\kilo\watt}. Under
the AIF fleet, the afternoon load remains flatter, and when the bottom floor
rises at 15:00 the middle floor stays at about \SI{20}{\kilo\watt} instead of
following it, which limits the building peak to \SI{83.8}{\kilo\watt}. It
should also be noted that the AIF fleet starts conditioning earlier in the
morning, which appears as a short spike at around 04:30.

\subsection{Deterministic Year and First-Year Behaviour}

Table~\ref{tab:w0} reports the results on the deterministic year, which is also
the year on which the learning controllers were trained. Here, the AIF fleet
reaches \SI{84.7}{\kilo\watt} and a CF of 0.876, and it already outperforms
every baseline in its first year of operation, starting from the physical prior
(\SI{86.0}{\kilo\watt}, CF of 0.855). Two further results of this table are
worth mentioning. First, the proactive RBC violates the deadband in 10.5\,\% of
the zone steps, since in winter some core zones receive identical heating and
cooling setpoints. Second, even without the congestion signal, learning the
transition model alone lowers the peak of the AIF fleet from
\SI{100.1}{\kilo\watt} in its first year to \SI{92.9}{\kilo\watt}.

\begin{table}[t]
\caption{Deterministic mixed year (W0). AIF: mean of the last three of five
adaptation years. DB: deadband violations (\% of zone steps).}
\label{tab:w0}
\centering
\footnotesize
\setlength{\tabcolsep}{3pt}
\begin{tabular}{lcccccc}
\toprule
Controller & Peak & CF & CS$_{occ}$ & DB & MWh & k\$ \\
\midrule
RBC reactive            & 93.2  & 0.987 & 0.918 & 0.0  & 103.8 & 52.2 \\
RBC proactive           & 92.4  & 0.980 & 0.947 & 10.5 & 137.2 & 60.2 \\
RBC demand-limiting     & 92.4  & 0.979 & 0.947 & 9.7  & 136.0 & 60.1 \\
MADDPG cold             & 104.7 & 0.997 & 0.928 & 1.6  & 118.6 & 55.3 \\
MADDPG warm (reactive)  & 99.0  & 0.958 & 0.934 & 1.6  & 119.1 & 55.8 \\
MADDPG + floor meters   & 91.9  & 0.913 & 0.930 & 1.9  & 117.4 & 54.6 \\
AIF, no congestion      & 92.9  & 0.974 & 0.939 & 0.0  & 119.4 & 56.6 \\
AIF (running reference) & 84.8  & 0.880 & 0.917 & 0.0  & 114.3 & 53.6 \\
AIF (fixed reference)   & 84.7  & 0.876 & 0.923 & 0.0  & 115.4 & 54.1 \\
\bottomrule
\end{tabular}
\end{table}

These numbers are better than those obtained on W1, because the learning
controllers are evaluated on the year they were trained on; for this reason, we
regard W1 as our main result. In addition, the strong first year does not fully
carry over to the noisy years: an AIF fleet starting from scratch on a noisy
year beat the first-year peak of the AIF baseline in four of five years with
the fixed reference (and in three of five with the running reference), although
it was cheaper in all five.

\subsection{Unseen Climates}

Table~\ref{tab:w2} presents the results on the hot and cool climates. In the
cool climate, the AIF fleet matches the peak of the proactive RBC
(\SI{48.5}{\kilo\watt}) while being \$4.6k cheaper. In the hot climate,
however, it performs poorly: its comfort drops to a $\mathrm{CS}_{occ}$ of
0.798, the lowest among all controllers, and its peak is higher than that of
both main RBCs. When the cooling plant operates at full capacity for long
periods, the only way to reduce the peak is to sacrifice a considerable amount
of comfort. Allowing the agents to keep learning online recovers a large part
of the comfort ($\mathrm{CS}_{occ}$ of 0.846, against 0.798 when frozen), but
the peak then rises to \SI{115.6}{\kilo\watt}. Moreover, in this climate the
fixed calendar of the summer comfort band penalises every controller, since on
1~October the band switches to the winter values while the outdoor temperature
is still high.

\begin{table}[t]
\caption{Unseen climates (W2), one year each.}
\label{tab:w2}
\centering
\footnotesize
\setlength{\tabcolsep}{3pt}
\begin{tabular}{lcccccc}
\toprule
 & \multicolumn{3}{c}{Hot} & \multicolumn{3}{c}{Cool} \\
\cmidrule(lr){2-4}\cmidrule(lr){5-7}
Controller & kW & CS$_{occ}$ & k\$ & kW & CS$_{occ}$ & k\$ \\
\midrule
AIF (fixed reference)   & 102.0 & 0.798 & 80.1 & 48.5 & 0.928 & 27.9 \\
AIF, no congestion      & 103.8 & 0.874 & 83.1 & 55.2 & 0.935 & 28.9 \\
RBC reactive            & 99.9  & 0.840 & 78.7 & 50.5 & 0.921 & 25.6 \\
RBC proactive           & 95.2  & 0.884 & 80.6 & 48.5 & 0.947 & 32.5 \\
RBC demand-limiting     & 110.0 & 0.884 & 81.9 & 48.5 & 0.947 & 32.2 \\
MADDPG cold             & 109.4 & 0.848 & 82.2 & 61.9 & 0.906 & 28.1 \\
MADDPG warm (reactive)  & 107.8 & 0.860 & 82.8 & 61.5 & 0.922 & 28.5 \\
MADDPG + floor meters   & 99.1  & 0.848 & 81.2 & 57.7 & 0.899 & 27.6 \\
\bottomrule
\end{tabular}
\end{table}

\subsection{AIF component contribution}
\label{sec:mechanism}

\begin{table}[t]
\caption{AIF ablations and mechanism controls on the deterministic year (mean
of the last three adaptation years). All rows use the running reference
unless stated otherwise.}
\label{tab:ablation}
\centering
\footnotesize
\setlength{\tabcolsep}{2pt}
\begin{tabular}{lccc}
\toprule
Configuration & Peak (kW) & CF & CS$_{occ}$ \\
\midrule
no congestion, fixed weights       & 92.9 & 0.974 & 0.939 \\
adaptive weights only              & 92.9 & 0.976 & 0.939 \\
congestion 5 only                  & 92.3 & 0.934 & 0.930 \\
congestion 4 + adaptive weights    & 92.0 & 0.923 & 0.924 \\
\textbf{congestion 5 + adaptive weights} & \textbf{84.8} & \textbf{0.880} & 0.917 \\
\quad fixed reference              & 84.7 & 0.876 & 0.923 \\
\quad earlier inference code       & 84.8 & 0.880 & 0.921 \\
\quad horizon 4\,/\,16 steps       & 87.1\,/\,85.0 & 0.891\,/\,0.929 & 0.919\,/\,0.904 \\
\quad stochastic actions           & 89.7 & 0.901 & 0.927 \\
\midrule
\multicolumn{4}{l}{\emph{Signal source, congestion weight 5\,/\,8:}} \\
\quad own floor                    & 91.9\,/\,85.0 & 0.919\,/\,0.877 & 0.921\,/\,0.893 \\
\quad whole building               & 92.6\,/\,83.7 & 0.928\,/\,0.874 & 0.921\,/\,0.892 \\
\quad 24\,h old                    & 92.8\,/\,92.9 & 0.931\,/\,0.963 & 0.926\,/\,0.912 \\
\quad constant                     & 92.9\,/\,92.9 & 0.961\,/\,0.959 & 0.933\,/\,0.930 \\
\bottomrule
\end{tabular}
\end{table}

Table~\ref{tab:ablation} isolates the contribution of each component of the AIF
controller on the deterministic year. As it is shown,
neither the congestion signal nor the adaptive energy weights lower the peak on
their own; only their combination does, and only from a congestion weight of 5
onwards, since at a weight of 4 the peak remains at \SI{92}{\kilo\watt}. The
peak therefore behaves like a switch rather than a slope, which is consistent
with the coarse action set: on the design day, the fleet either keeps its
cooling setpoint or gives up one whole step.

It is also observed that a constant signal with the same mean value,
or the same signal delayed by 24 hours, never reaches the low-peak regime, even
with a congestion weight of 8. Consequently, the effect is neither a higher
energy price in disguise nor a learned daily pattern; instead, the agents react
to what the building is drawing at that moment. Signals derived from the
agent's own floor or from the whole building also work, but they require a
weight of 8 and cost about 0.03 more in comfort, whereas the signal from the
other floors reaches the same peak at a lower comfort cost.

Since every agent of a floor
receives the same congestion signal, agents with identical weights react in
lockstep. When we replace the adaptive weights with fixed random weights per
agent, the fleet reaches the low-peak regime in 4 of 8 random draws, whereas
the adaptive weights find a working combination every time, without anyone
having to choose them. In practice, the weights rarely settle: most of the time
each agent sits at one of the two bounds and switches frequently, so that at any
given moment some agents save energy while others protect comfort.

Providing the
MADDPG actors with the three floor meters lowered their peak by
\SI{9.0}{\kilo\watt} and their CF by 0.057 in every noisy year, while slightly
improving comfort. This result independently supports the explanation above,
namely that live load information is what allows independent agents to spread
their demand. Given the same information, the AIF fleet still achieves a lower
peak (by \SI{8.7}{\kilo\watt}), a lower cost and a lower energy consumption in
every year, but its CF advantage shrinks to 0.007, which lies within the
year-to-year variability. It should also be noted that MADDPG required 50
training years and a centralised critic to reach this performance, whereas the
AIF fleet adapted for five years without any centralised training.

Furthermore, the additional signal did not change what MADDPG learned in terms
of the metrics tracked during training: both cold-start runs reach the same
final reward and comfort, and their comfort stops improving after about 30 to
40 training years, while the shaped reward still increases slowly at year 50.

\subsection{Deployment on Edge Hardware}
\label{sec:latency}

For the deployment we used the open-source
\texttt{wactorz} actor runtime where each zone agent runs as an independent process and communicates only over MQTT, subscribing
to the observation stream and publishing the setpoints of its zone. 
Decision latency was measured with all 15 AIF agents planning concurrently on a
Raspberry~Pi~5 while driving the simulator over MQTT. An instrumented bridge
timestamps the full observation-to-action round trip, so the reported figures
include the messaging overhead rather than computation alone. Over 90 timed
decisions (after discarding 10 warm-up decisions), we measured a mean latency of
\SI{268}{\milli\second}, a median of \SI{238}{\milli\second}, a 95th percentile
of \SI{442}{\milli\second} and a maximum of \SI{822}{\milli\second}, which is
negligible compared with the 15-minute control step.

\section{Discussion and Limitations}
\label{sec:discussion}

\begin{figure}[t]
\centering
\includegraphics[width=0.8\columnwidth]{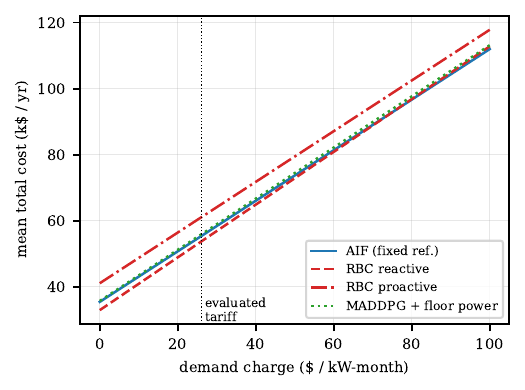}
\caption{Mean yearly cost on the noisy test years as a function of the demand
charge. The dotted line marks the evaluated tariff (\$26/kW-month).}
\label{fig:tariff}
\end{figure}

\textbf{The AIF fleet trades energy for demand.} Compared with the reactive RBC
at similar comfort, the AIF fleet lowers the sum of monthly peaks by
\SI{33}{\kilo\watt} per year and spreads the demand of the three floors, but it
consumes 11~MWh more energy. Whether this trade-off pays off depends on the
tariff: with the demand charge of \$26/kW-month that we evaluated, it does not,
and the fleet would break even with the reactive RBC only at a demand charge of
about \$75/kW-month (Fig.~\ref{fig:tariff}). Nevertheless, lower and
better-staggered demand can also be valuable beyond the electricity bill, for
instance under coincident-peak charges or local grid constraints, which our
tariff does not price.

\textbf{A well-tuned RBC is a strong baseline.} On this building, the proactive
RBC is the most comfortable controller, while the reactive RBC is the cheapest,
and neither learning approach outperforms both of them on comfort and cost
simultaneously. What the AIF fleet adds is lower and better-staggered demand,
a single setting (the congestion weight) that covers the range between the two
RBCs, and a decision mechanism that can be inspected.

\textbf{Single-year peaks are fragile.} Since an annual peak is a single
15-minute value, results obtained on a single year can be misleading: on the
deterministic year, one RBC setting appeared better than the AIF fleet, and
this advantage disappeared on the noisy years. Testing on several noisy years,
comparing controllers year by year, and reporting the sum of monthly peaks
changed several of our conclusions, and we recommend the same practice for HVAC
benchmarks.

\textbf{Scope and limitations.} Our study has the following limitations that should
be taken into account when interpreting the results:
\begin{itemize}
\item We evaluated a single building and a single tariff; the noisy years vary
the weather, but not the building or the occupancy patterns.
\item The AIF fleet is the least comfortable of the strong controllers, and in
a hot climate its comfort degrades considerably when it is not allowed to
adapt.
\item Comfort is measured on weekdays between 07:00 and 19:59, although
EnergyPlus also reports occupants outside these hours, for example in the early
morning, in the evening and on Saturdays.
\item Finally, each MADDPG variant was trained once.
\end{itemize}

\section{Conclusion}
\label{sec:conclusion}

In this paper, we presented a decentralised controller for multi-zone HVAC
systems in which fifteen active-inference agents, one per zone and exchanging no
messages, coordinate the demand of a building that shares one air loop per
floor. Evaluated on five noisy weather years that were not seen during
training, the AIF fleet achieved the lowest mean annual peak among all the
controllers that we tested and a lower coincidence factor than any baseline.
Our analysis showed that this
coordination stems from a live load signal combined with agents that weigh
energy differently, and that a deep reinforcement learning controller given the
same meters improves for the same reason. The price of this behaviour is an
energy consumption about 10\,\% higher than that of a simple reactive rule,
which pays off only where demand is charged heavily. Finally, we showed that
all 15 agents run concurrently on a Raspberry~Pi~5 with a p95 latency of
\SI{442}{\milli\second}, measured over the full MQTT round trip.

Future work follows directly from the limitations of this study. In particular,
we plan to gate the congestion signal by comfort, so that the agents do not
sacrifice comfort in hot climates, to refine the action set, and to evaluate
the approach on more buildings and tariffs. Moreover, seeding the transition
model with the structural knowledge that makes the proactive RBC so comfortable
is a promising direction for combining the strengths of both approaches.

\section*{Acknowledgment}
This work was funded by the European Network of Excellence dAIEDGE under Grant Agreement No 101120726.

\bibliographystyle{IEEEtran}
\bibliography{refs}

\end{document}